\documentclass{aa}
 \usepackage{graphicx} 
\begin{document}

   \title{Vertical abundance stratification in the blue horizontal branch star HD~135485\thanks{Based
on observations made with ESPaDOnS at the Canada-France-Hawaii Telescope (CFHT) operated by the
National Research Council (NRC) of Canada, the Institut des Sciences de l'Univers of the Centre National
de la Recherche Scientifique (CNRS) and the University of Hawaii and on observations made with Echelle
Spectrograph on the McDonald Observatory 2.1-meter Otto Struve Telescope.}}

   \author{V.R. Khalack \inst{1,}\inst{2}, F. LeBlanc\inst{1}, D. Bohlender\inst{3}, 
    G.A. Wade\inst{2}, B.B. Behr\inst{4}\thanks{Current address:
US Naval Observatory, 3450 Massachusetts Avenue NW, Washington DC 20392}}

   \offprints{V. Khalack \\ \email{khalakv@umoncton.ca}}

   \institute{ D\'epartement de Physique et d'Astronomie,
               Universit\'e de Moncton,
               Moncton, N.-B., Canada E1A 3E9
               \and
               Department of Physics, Royal Military College of Canada,
               PO Box 17000 stn `FORCES', Kingston, Ontario, Canada K7K 4B4
               \and
               National Research Council of Canada,
               Herzberg Institute of Astrophysics,
               5071 West Saanich Road, Victoria, BC, Canada V9E 2E7
               \and
               Department of Astronomy, University of Texas at Austin, 1 University
               Station C1400, Austin TX 78712-0259, USA
             }

   \date{Received {\it date will be inserted by the editor}\
         Accepted {\it date will be inserted by the editor}
        }

\abstract 
{It is commonly believed that the observed
overabundances of many chemical species relative to the expected cluster
metallicity in blue horizontal branch (BHB) stars appear as a result
of atomic diffusion in the photosphere. The slow rotation of BHB stars
(with $T_{\rm eff}>$ 11,500K), typically $v\sin{i}<$ 10 km s$^{-1}$,
is consistent with this idea.}
{In this work we search for observational evidence of vertical chemical
stratification in the atmosphere of HD~135485. If this evidence
exists, it will demonstrate the importance of atomic diffusion processes in
the atmospheres of BHB stars.}
{We undertake an extensive abundance stratification analysis of the
atmosphere of HD 135485, based on recently acquired high resolution
and S/N CFHT ESPaDOnS spectra and a McDonald-CE spectrum.}
{Our numerical simulations show that nitrogen and sulfur reveal
signatures of vertical abundance stratification in the stellar atmosphere. It
appears that the abundances of these elements increase toward the upper atmosphere.
This fact cannot be explained by the influence of microturbulent velocity,
because oxygen, carbon, neon, argon, titanium and chromium do not show similar behavior
and their abundances remain constant throughout the atmosphere. It seems
that the iron abundance may increase marginally toward the lower atmosphere.
This is the first demonstration of vertical
abundance stratification of metals in a BHB star.}
{}


\titlerunning{Vertical abundance stratification in HD~135485}
\authorrunning{Khalack et al.}

\maketitle

\section{Introduction}

HD~135485 appears to be an evolved star that lies close to or on the
horizontal branch (Trundle et al. \cite{Trundle+01}). It has
mid-B spectral type and shows many prominent absorption lines of
metallic species in its spectrum. The most recent absolute and
differential abundance analysis of HD~135485 was performed by
Trundle et al. (\cite{Trundle+01}). They found a large
(0.5-1.0 dex) enhancement of metal
abundances in comparison with the solar composition. Similar
results were obtained previously by Sch\"{o}nberner
(\cite{Schonbe73}) and Dufton (\cite{Dufton73}) from analysis of
moderate-resolution photographic spectra. In an attempt to explain the
anomalous chemical composition in the atmosphere of HD~135485,
Dufton (\cite{Dufton73}) considered the possibility of binary system
evolution, where the mass flow onto HD~135485 from a high mass
component may have created the observed enhancements of metal
species. In order
to try to prove this hypothesis, Dufton investigated the
radial velocity during a five month period, but
found no variation (see Trundle et al. (\cite{Trundle+01}) for details).

Another possible explanation proposed for the
unusually high metallicity is that the star was formed in
a region of high metallicity, such as the Galactic Center, and during
the long period of its stellar evolution it could have traveled over a
large distance ($\sim$26 kpc) as demonstrated by Trundle et al.
(\cite{Trundle+01}) from analysis of the kinematical properties
of HD 135485.

\begin{table*}[th]
\parbox[t]{\textwidth}{
\caption[]{Journal of spectroscopic observations of HD~135485.}
\begin{tabular}{ccccccc}
\hline
\hline
 Date  &     HJD    &Instrument&$t_{\rm exp}$& S/N ratio & Coverage & Resolution\\
 (UT)  &(2450 000 +)& & (s)   &(pixel)$^{-1}$& $\lambda$ (\AA) &  \\
\hline
2001 Mar 09& 1978.9242 &    CE    & 600 &  67 & 4760 -  5570 & 60000 \\
2005 May 19& 3509.8677 & ESPaDOnS & 600 & 150 & 3720 - 10290 & 80000 \\
2005 May 19& 3509.8753 & ESPaDOnS & 600 & 140 & 3720 - 10290 & 80000 \\
\hline
\end{tabular}
}
\label{tab1}
\end{table*}

Together with general enrichment of all metals (except nickel which is significantly
depleted) with respect to hydrogen, Trundle et al. (\cite{Trundle+01}) have also found
a slightly enhanced helium abundance of $\sim$0.3 dex. They interpret this fact
along with the anomalously high ($\sim$1.3 dex) nitrogen overabundance as an indication
that the carbon-nitrogen (CN) cycle took place during the hydrogen-burning phase on
the main sequence. The relatively normal carbon enhancement might reflect a balance
between its being destroyed by the CN cycle and created via helium core flashes
during the red giant phase (which could also explain the observed neon
enrichment in the atmosphere). These facts and the present position of HD~135485 on the
Hertzsprung-Russell (HR) diagram argue that this star is evolved and probably
belongs to the blue horizontal branch (BHB) stars. The observed low (projected) rotational velocity
($v\sin{i}<$ 4 km s$^{-1}$) of HD~135485, the observed chemical peculiarities (except
the helium enrichment) and the effective temperature $T_{\rm eff}=$ 15500K (Trundle et al.
\cite{Trundle+01}) are typical
for BHB stars.

Comprehensive surveys of BHB star abundances show that the stars hotter
than $T_{\rm eff} \simeq$ 11,500K have abundance anomalies when compared to
the BHB stars of lower effective temperature in the same
cluster\footnote{Most of the known BHB stars are found in globular clusters.}
(Glaspey et al. \cite{Glaspey+89}, Grundhal et al. \cite{Grundhal+99}).
Also, Peterson, Rood \& Crocker (\cite{Peterson+95}); Behr et al. (\cite{Behr+00})
and Recio-Blanco et al. (\cite{RB+04})
showed the existence of a discontinuity in
stellar rotation velocity distribution of BHB stars at $T_{\rm eff} \simeq $ 11,500K.
All of the hotter stars show modest rotation ($v\sin{i}<$ 10 km s$^{-1}$),
while the cooler stars are rotating more rapidly.

Other observational results show both
photometric jumps (Grundhal et al. \cite{Grundhal+99}) and photometric gaps
(Ferraro et al. \cite{Ferraro+98}), also occurring at $T_{\rm eff} \simeq $ 11,500K, as
well as
low measured surface gravities. 
The low gravity, abundance anomalies and slow rotation suggest that microscopic
atomic diffusion is effective in stellar atmospheres of the BHB stars with
$T_{\rm eff} \geq$ 11,500K. 
In this picture, the competition between radiative acceleration and
gravitational settling yields a net acceleration on atoms, which results
in their diffusion through the atmosphere.
A consequence of this process
is vertically-stratified abundances of chemical elements.
The aim of this paper is to attempt to detect observationally
such vertical stratification of elements in the atmospheres of BHB stars. Direct measurement of
this stratification from line profile
analysis would provide a convincing argument in favour of efficient chemical diffusion
in the atmospheres of BHB stars with $T_{\rm eff} \geq$ 11,500K.

The most important difference between HD~135485 and most BHB stars is that it has
an enhanced helium abundance. Therefore HD~135485 might be a very
young BHB star in which helium has not yet settled. However, a detailed investigation of this
possibility is outside the
scope of this paper.

To verify if the abundances of some chemical species are vertically
stratified in the atmosphere of HD~135485, we undertook a detailed analysis
of available
high-resolution spectra employing a modification of the approach for abundance stratification
analysis developed by Ryabchikova et al. (\cite{Ryabchikova+05}).
In Sect.~\ref{obs} we discuss the properties of the observed spectra, and in
Sect.~\ref{mod} we describe the atmospheric parameters of HD~135485 and modelling details.
The results of abundance determinations are considered in Sect.~\ref{res},
while a description of vertical stratification for some chemical species
is shown in Sect.~\ref{vert}. A discussion of our results follows in Sect.~\ref{discuss}.

\section{Observations}
\label{obs}

Spectroscopic observations of HD~135485 were undertaken in March 2001 and
May 2005. The journal of spectroscopic observations is shown in
Table~\ref{tab1}, where individual columns give the UT date of the observation,
heliocentric Julian Date of 
the observation, the instrument specification,
exposure time, signal-to-noise (S/N) ratios per pixel
(the peak typically occurs around 5400 \AA), spectral coverage and resolving power
(which depends on central wavelength).

\subsection{McDonald Cassegrain Echelle spectrum}

The first spectrum of HD~135485 (from Table~\ref{tab1}) was obtained in the
framework of a study of the rotational characteristics ($v\sin{i}$) of BHB stars
(Behr~\cite{Behr03}) using
the Cassegrain Echelle (CE) Spectrograph (McCarthy et al.~\cite{McCarthy+93})
on the McDonald Observatory 2.1-meter Otto Struve Telescope. This instrument
provides a nominal resolving power of $R\sim 60000$.
The signal-to-noise ratio (per pixel) was computed from
the reduced spectrum by smoothing the spectral regions identified as
continuum, and then computing the rms deviation between the observed
and smoothed spectra.

The package of routines developed by McCarthy (\cite{McCarthy90}) for the FIGARO
data analysis package (Shortridge~\cite{Shortridge93}) was employed to reduce
the spectrum. The detailed description of the reduction procedure for this particular
spectrum, including the wavelength calibration routine and the continuum normalization,
is provided by Behr~(\cite{Behr03}). The procedure follows a standard
prescription, with bias subtraction, flat-fielding, order extraction,
and wavelength calibration from thorium-argon arc lamp observations.
Cosmic ray hits were identified and removed by hand for maximum
spectral quality. No smoothing was applied to the spectra.

\subsection{CFHT-ESPaDOnS spectra}

The other two spectra were obtained using the new ESPaDOnS (Echelle SpectroPolarimetric
Device for Observations of Stars) spectropolarimeter at the Canada-France-Hawaii Telescope
(CFHT). ESPaDOnS is fundamentally similar in construction to the
MuSiCoS spectropolarimeter (Donati et al. 1999) and allows the acquisition of an essentially continuous
spectrum throughout the spectral range 3700 \AA\, to 10500 \AA\, in a single exposure.
The optical characteristics of the spectrograph as well as the spectropolarimeter
observing procedures, are described by Donati et al. (in preparation)\footnote{
For more details about this instrument, the reader is invited to visit
{\rm www.ast.obs-mip.fr/projects/espadons/espadons.html}}.
Observations were performed in the spectroscopic `sky only' mode of the instrument that provides
the highest resolving power ($R\sim80000$).
This resolution corresponds to velocity elements of 3.75 km s$^{-1}$.

The ESPaDOnS spectra were reduced using the Libre-ESpRIT reduction tool
(Donati et al., in preparation), which is the most recent release of ESpRIT
(Donati et al., \cite{Donati+97}). The continuum of each spectral order was
normalized by fitting a 3rd to 5th order polynomial (as selected by eye)
to regions free of prominent spectral lines and not located immediately at
the beginning or at the end of the spectral order.

\section{Line profile simulations}
\label{mod}

\subsection{Stellar atmosphere parameters}

The line profile simulations were performed 
using a Phoenix (Hauschildt et al.~\cite{Hauschildt+97})
LTE (Local Thermodynamic Equilibrium) stellar atmosphere
model with enhanced (+1.0 dex) metallicity.
The model 
is calculated assuming $T_{\rm eff}=15,500$K, and $\log{g}=4.0$,
which according to Trundle et al. (\cite{Trundle+01}) provide the
best description of observable line profiles of HD~135485.
Those authors adopted a microturbulent velocity of 5 km s$^{-1}$ to
provide a constant behavior of abundances over the range of
equivalent widths measured for sulfur and nitrogen ions. Behr (\cite{Behr03}) has adopted for
this star $\xi=$ 1.4 km s$^{-1}$ and obtained $v\sin{i}=$ 0 km s$^{-1}$.

From our perspective, the physical motivation for adopting a microturbulent velocity is
unclear. According to Landstreet (1998), photospheric convection should be very
weak at the effective temperatures of BHB stars. Landstreet
found that spectroscopically-derived microtubulence had decreased from around 4~km s$^{-1}$ in late
A stars with $T_{\rm eff}\sim 8000$~K to 0.5~km s$^{-1}$ in early A stars with $T_{\rm eff}\sim 10500$~K.
In particular, strong microtubulence (e.g. 5~km s$^{-1}$ as adopted by Trundle et al. \cite{Trundle+01})
should be accompanied by strong line asymmetries, which we do not observe.
We therefore realistically expect no detectable microturbulence in HD~135485. Of course, Landstreet's analysis was
performed for main sequence stars; we might reasonably expect that BHB stars show some differences, although the
atmospheric parameters of HD~135485 are very similar to those of main sequence stars.


Based on this guidance, we have tentatively adopted a microturbulent velocity 0 km s$^{-1}$ for our abundance
analysis of HD~135485. Using zero microturbulence in spectral synthesis we estimated stellar rotational
velocity $v\sin{i}=$2 - 3 km s$^{-1}$.
This is an important assumption for several reasons. First, because diffusion is a slow
process, the mixing which might be implied by the presence of microturbulence could serve to reduce or erase chemical
stratification. Also, adoption of microturbulence modifies line strengths, leading to systematic changes in inferred mean
abundances as well as abundance trends with line strength. As we shall see in Sect. 5, the validity of the assumption of
zero microturbulence can be tested, and it is fully consistent with our spectroscopic observations.

\subsection{Procedure}
\label{proc}

All of the available spectra were examined to compose a line list
for chemical
species that are suitable for abundance and stratification analysis.
The line identification is performed using the VALD-2
(Kupka~et~al.~\cite{Kupka+99}; Ryabchikova~et~al.~\cite{Ryab+99}) and
NIST\footnote{\rm http://physics.nist.gov/PhysRefData/ASD/index.html}
(version 3.0.3) line databases (Reader et al.~\cite{Reader+02}).
The same sources are used to specify the atomic data for the
selected lines. For Fe\,{\sc ii} and Cr\,{\sc ii} lines we extracted atomic data
from Raassen \& Uylings\footnote{\rm ftp://ftp.wins.uva.nl/pub/orth} (\cite{RU+98}),
while for some Fe\,{\sc ii} lines we also used the atomic data recently obtained
by Fuhr \& Wiese (\cite{F+W06}).
An example of the first ten lines from the line list selected for the abundance analysis
of N\,{\sc ii} lines is shown in Table~\ref{tab2}\footnote{Table~\ref{tab2} is presented
full in the electronic edition of Astronomy and Astrophysics.
A portion is shown here for guidance and content.},
where references identify sources for the adopted $\log gf$ values, while the given energy
levels are taken mostly from the NIST database.

\begin{table}[t]
\centering
\caption[]{List of spectral lines used for the abundance analysis.}
\begin{tabular}{crccl}
\hline
\hline
\multicolumn{5}{c}{ N\,{\sc ii}} \\
\hline
$\lambda$, \AA&$\log gf$& $E_i, cm^{-1}$&$\log \gamma_{\rm rad}$& Ref.\\
\hline
4447.030& 0.228& 164610.76& 9.16 & NIST \\
4507.556&-0.817& 166678.64& 9.33 & VALD \\
4601.478&-0.428& 148940.17& 9.23 & VALD \\
4613.868&-0.665& 148940.17& 9.23 & VALD \\
4621.396&-0.514& 148940.17& 9.23 & VALD \\
4630.539& 0.094& 149076.52& 9.15 & VALD \\
4779.722&-0.587& 166521.69& 9.58 & VALD \\
4803.287&-0.113& 166678.64& 9.58 & VALD \\
4987.376&-0.555& 168892.21& 9.34 & VALD \\
4994.367&-0.069& 168892.21& 9.34 & VALD \\
\hline
\end{tabular}
\label{tab2}
\end{table}

All line profiles are simulated using the {\sc Zeeman2} spectrum
synthesis code (Landstreet~\cite{Landstreet88};
Wade~et~al.~\cite{Wade+01}) assuming Gaussian instrumental
profiles. The stability of the ESPaDOnS instrumental profile has
been investigated using the calibration spectra of a ThAr quartz
lamp. We find the instrumental profile to be stable during the night
of observation and to reveal no dependence on the intensity of the line
profile. The {\sc Zeeman2} code has been modified (Khalack~\&~Wade \cite{Khalack+Wade06}) to
allow for an automatic minimization of the model parameters using the
{\it downhill simplex method} (Press~et~al.~\cite{press+}).
The relatively poor efficiency of the downhill simplex method, requiring a
large number of function evaluations, is a well known problem. Repeating the minimization
routine several times in the vicinity of a supposed minimum in the parameter space allows
us to check if the method converges to the global minimum.

To search for the presence of abundance stratification, we have
used two different methods. In the first method, we estimate the
abundance of a chemical element from independent analysis of each line
profile. For every layer of the stellar atmosphere model (it contains 50 layers)
we calculate the line optical depth $\tau_{\rm \ell}$ in the line core.
We suppose that each analyzed profile is formed mainly at $\tau_{\rm \ell}$=1,
which corresponds to a particular layer of the stellar atmosphere. Finally,
for this layer the respective continuum optical depth
$\tau_{\rm 5000}$ is specified. In this way, we have built the
scale of optical
depths $\tau_{\rm 5000}$ to track vertical abundance stratification through the analysis
of all the available line profiles for a chosen chemical element.

Three free model parameters (the element's abundance,
line radial velocity $V_{\rm r}$ and $v\sin{i}$) were derived from each line profile
using the aforementioned automatic minimization routine.
HD~135485 has very sharp absorption lines and even for resolution R=80000
a line profile usually consists of only 8 to 15 bins.
This leads to the fact that the downhill simplex method fails to converge
to the global minimum in approximately 2 to 5\% of analyzed
line profiles. This means that the automatic minimization routine remains in the vicinity of
a local minimum and results in a higher final value of the $\chi^2$-function.
Repeating the minimization routine with different starting values of the model parameters
usually solves this problem and provides a better fit, which is checked using a
few additional runs of the simulation routine.
Sometimes, analysis of a line profile results in a
radial velocity which differs significantly from the average $V_{\rm r}$=4.2 km s$^{-1}$ (obtained
from analysis of all line profiles of the same chemical element and 
other chemical species). This may be evidence for line
misidentifications or inaccurate line wavelengths.
Therefore lines with radial velocities which differ more than 1.5 km s$^{-1}$
from the average are not taken into account in the modeling
and are not included in Table~\ref{tab2}.

In the second method, we only analyzed those lines which appear to be good candidates in the
previous approach and consider the two-step stratified abundance model with a linear transition zone
(Ryabchikova et al. \cite{Ryabchikova+03}; Wade et al. \cite{Wade+03})
to describe the abundance distribution. In this case, all the selected lines were analyzed
simultaneously and in the minimization routine we operated with 6 free model parameters:
the element's abundance
in the deeper layers, the (standard) optical depth $\tau_{\rm 1}$ of the transition zone lower (deeper) boundary
(where the element's abundance begins its linear increase or decrease), the optical depth $\tau_{\rm 2}$ of the
transition zone
upper boundary, the difference between the abundances at two boundaries, the stellar radial velocity and the $v\sin{i}$.
We note that for the second method the downhill simplex method never failed to reach the global minimum.
This means that information stored in all of the analyzed lines, modeled simultaneously, is statistically sufficient to
provide a unique solution.
To determine the averaged abundances for all of the analyzed chemical species,
we simultaneously fit the available line profiles using the same three free model parameters
used in the first approach.

\begin{table}[t]
\centering
\caption[]{Mean atmospheric abundances $\log(N/N_{\rm tot})$ of HD~135485
with uncertainties equal to the standard deviation of $n$
measured lines.}
\begin{tabular}{lccccc}
\hline
\hline
 &\multicolumn{2}{c}{CFHT-ESPaDOnS}& \multicolumn{2}{c}{McDonald-CE}& Sun \\
Ion  & $\log(N/N_{\rm tot})$& n  & $\log(N/N_{\rm tot})$& n  &$\log(N/N_{\rm tot})$ \\
\hline
He\,{\sc i}  &-0.58$\pm$0.16& 9 &-0.54$\pm$0.15& 3 & -1.11 \\
&&&&&\\
C\,{\sc i}   &-2.78$\pm$0.17& 10&    &   & -3.65 \\
C\,{\sc ii}  &-2.69$\pm$0.14& 15&-2.88$\pm$0.16& 4 & -3.65 \\
&&&&&\\
N\,{\sc i}   &-2.60$\pm$0.20& 3 &    &   & -4.26 \\
N\,{\sc ii}  &-3.00$\pm$0.17& 21&-3.01$\pm$0.18& 9 & -4.26 \\
&&&&&\\
O\,{\sc i}   &-2.82$\pm$0.13& 11&-2.74$\pm$0.19& 2 & -3.38 \\
O\,{\sc ii}  &-3.28$\pm$0.22& 9 &    &   & -3.38 \\
&&&&&\\
Ne\,{\sc i}  &-3.22$\pm$0.06& 4 &    &   & -4.20 \\
&&&&&\\
Na\,{\sc i}  &-4.22$\pm$0.24& 2 &    &   & -5.87 \\
&&&&&\\
Mg\,{\sc i}  &-3.62$\pm$0.20& 1 &    &   & -4.51 \\
Mg\,{\sc ii} &-3.73$\pm$0.12& 4 &    &   & -4.51 \\
&&&&&\\
Al\,{\sc ii} &-5.32$\pm$0.11& 7 &    &   & -5.67 \\
Al\,{\sc iii}&-5.10$\pm$0.09& 4 &    &   & -5.67 \\
&&&&&\\
Si\,{\sc ii} &-4.28$\pm$0.23& 9 &-4.23$\pm$0.11& 2 & -4.53 \\
Si\,{\sc iii}&-3.96$\pm$0.22& 2 &    &   & -4.53 \\
&&&&&\\
P\,{\sc ii}  &-5.81$\pm$0.12& 8 &-5.76$\pm$0.13& 3 & -6.68 \\
&&&&&\\
S\,{\sc ii}  &-4.43$\pm$0.18& 48&-4.47$\pm$0.17& 21& -4.90 \\
&&&&&\\
Cl\,{\sc ii} &-6.35$\pm$0.20& 1 &-6.47$\pm$0.20& 1 & -6.54 \\
&&&&&\\
Ar\,{\sc ii} &-5.03$\pm$0.05& 4 &-4.96$\pm$0.19& 2 & -5.86 \\
&&&&&\\
Ti\,{\sc ii} &-6.70$\pm$0.04& 5 &    &   & -7.14 \\
&&&&&\\
Cr\,{\sc ii} &-6.01$\pm$0.07& 5 &-5.89$\pm$0.14& 2 & -6.40 \\
&&&&&\\
Fe\,{\sc ii} &-4.08$\pm$0.16& 63&-4.16$\pm$0.18& 31& -4.59 \\
&&&&&\\
Sr\,{\sc ii} &-8.50$\pm$0.20& 1 &    &   & -9.12 \\
\hline
\end{tabular}
\label{tab3}
\end{table}

Initially, we specify arbitrarily the free model parameters in the range of
their possible values and simulate the line profiles. We then compare
the simulated profiles with the observed spectra, and calculate the reduced
$\chi^\mathrm{2}$, which we adopt as a measure of the fit
quality. The expression for the $\chi^\mathrm{2}$ function, which reflects the
agreement between the simulated profiles ($I_{\rm i,\lambda_j}$) and
observed spectra ($I^{obs}_{\rm i,\lambda_j}$) is given by:

\begin{equation} \label{chi2}
\chi^{2}_{I} = \displaystyle \frac{1}{N_{I}} \sum^{N_{I}}_{i=1}
\displaystyle \frac{1}{N_{i}} \sum^{N_{i}}_{j=1}
\left( \displaystyle \frac{I^{obs}_{i,\lambda_j}-I_{i,\lambda_j}}
{\sigma[I^{obs}_{i,\lambda_j}]}\right)^{2},
\end{equation}

\noindent where $\sigma[I^{obs}_{\rm i,\lambda_j}]$ corresponds to the measurement
errors, $N_{I}$ represents the number of profiles indexed by $i$, while
$N_{i}$ is the number of pixels in each analyzed line profile. In the first method
of the analysis we used $N_{I}$=1.
In fact, the simulated spectra are calculated with a resolution of 0.01 \AA\, and so do
not provide a direct coincidence of wavelengths in the simulated and observed spectra.
Therefore, during the $\chi^\mathrm{2}$ function evaluation the simulated spectral
intensity at the exact observed wavelength is calculated using a linear interpolation.

In order to evaluate the fit errors (and therefore the uncertainties on the
derived free parameters), we calculate deviations of the simulated profiles
produced as a result of small variations of each of the free parameters, thus
introducing a small shift along one axis in the $\chi^\mathrm{2}$ hyper-space
from the point of the function minimum value. Using this procedure, and taking
into account the uncertainties of the observational data and the obtained
minimum value of the $\chi^\mathrm{2}$-function, we can estimate the errors of the
best-fit parameters.

\section{Results of abundance analysis}
\label{res}

We have identified in the spectra of HD~135485 most of the lines
reported by Trundle et al. (\cite{Trundle+01}). 
No lines of Mn\,{\sc ii} and Ni\,{\sc ii} were detectable in
our spectra and these elements were therefore not analysed.
It seems that Trundle et al. (\cite{Trundle+01})
have determined the Mn\,{\sc ii} and Ni\,{\sc ii} abundances from the much stronger
ultraviolet lines of these species.  The
McDonald-CE spectrum has lower spectral resolution
than the ESPaDOnS data and is used mainly
to confirm the average abundance obtained with the ESPaDOnS spectra.

A preliminary LTE abundance analysis of helium lines suggests an enhanced
helium abundance in HD~135485, and the element also seems to be vertically stratified. The
helium overabundance obtained is in good agreement with the results reported by Trundle et al.
(\cite{Trundle+01}).

The mean photospheric abundances we derive for HD~135485 are reported in Table~\ref{tab3}.
The second and third columns contain respectively the mean abundance of the chemical species
and the number of analyzed line profiles in the ESPaDOnS spectra,
while the fourth and fifth columns represent
similar results obtained from analysis of the McDonald-CE spectrum.
The last column contains the solar atmospheric abundances
recalculated from $\log(N/N_{\rm H})$ data (Asplund et al. \cite{Asplund+05}).

\subsection{Light elements: C to Ar}
\label{light}

\begin{figure*}[th]
\parbox[t]{\textwidth}{
\centerline{%
\begin{tabular}{@{\hspace{+0.05in}}c@{\hspace{+0.1in}}c}
a) & b)\\
\includegraphics[angle=-90,width=3.5in]{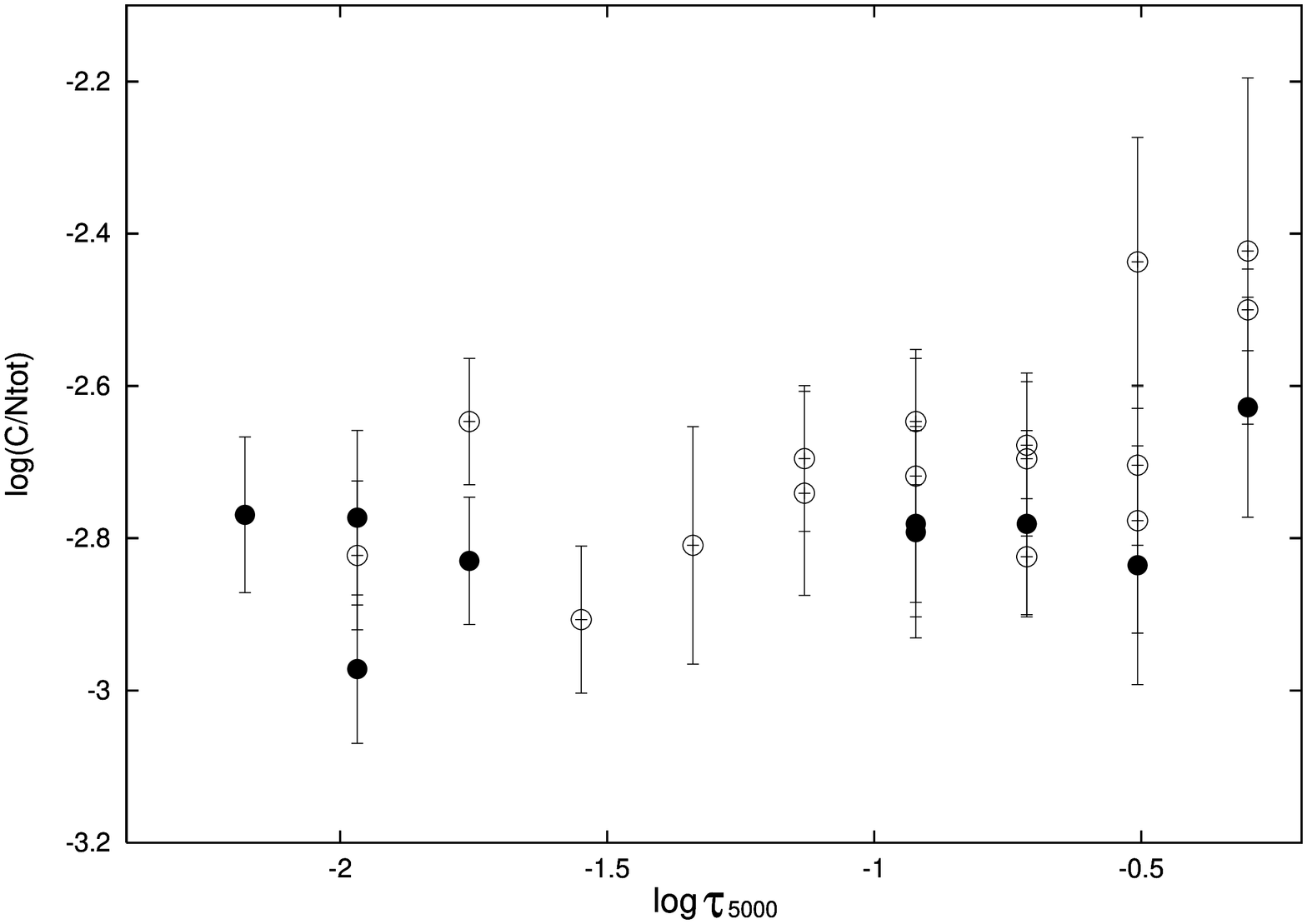} &
\includegraphics[angle=-90,width=3.5in]{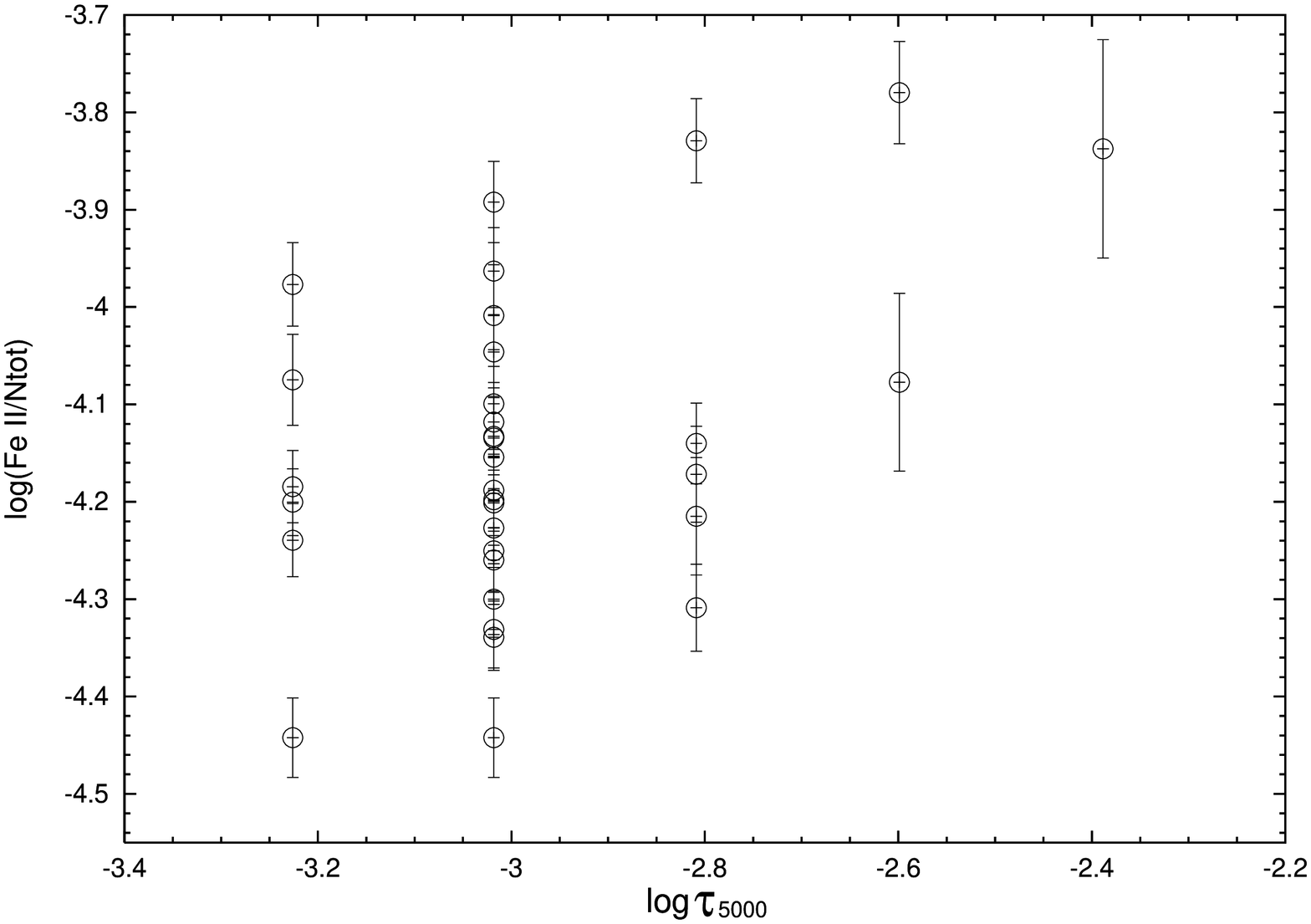}
\end{tabular}
}
\caption{Abundance estimates from the analysis of a) C\,{\sc i} (filled circles)
and C\,{\sc ii} (open circles) and b) Fe\,{\sc ii} (open circles) line profiles as
a function of line (core) formation optical depth. No {clear} signatures of
abundance stratification are visible.
The abundances obtained have approximately the same value for all optical depths at which
 carbon lines are formed.
\label{CFe}}
}
\end{figure*}

\begin{figure*}[th]
\parbox[t]{\textwidth}{
\centerline{%
\begin{tabular}{@{\hspace{+0.05in}}c@{\hspace{+0.1in}}c}
a) & b)\\
\includegraphics[angle=-90,width=3.5in]{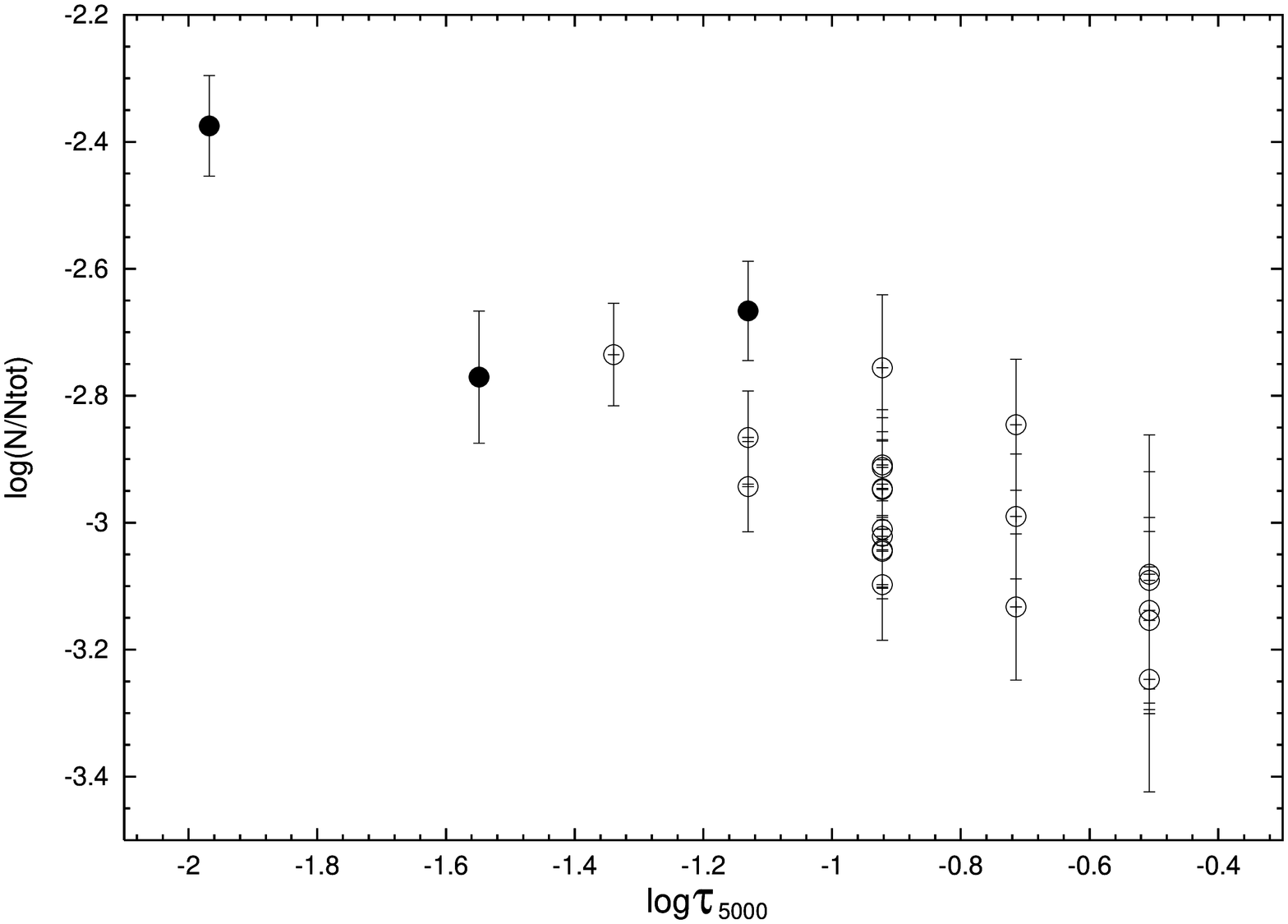} &
\includegraphics[angle=-90,width=3.5in]{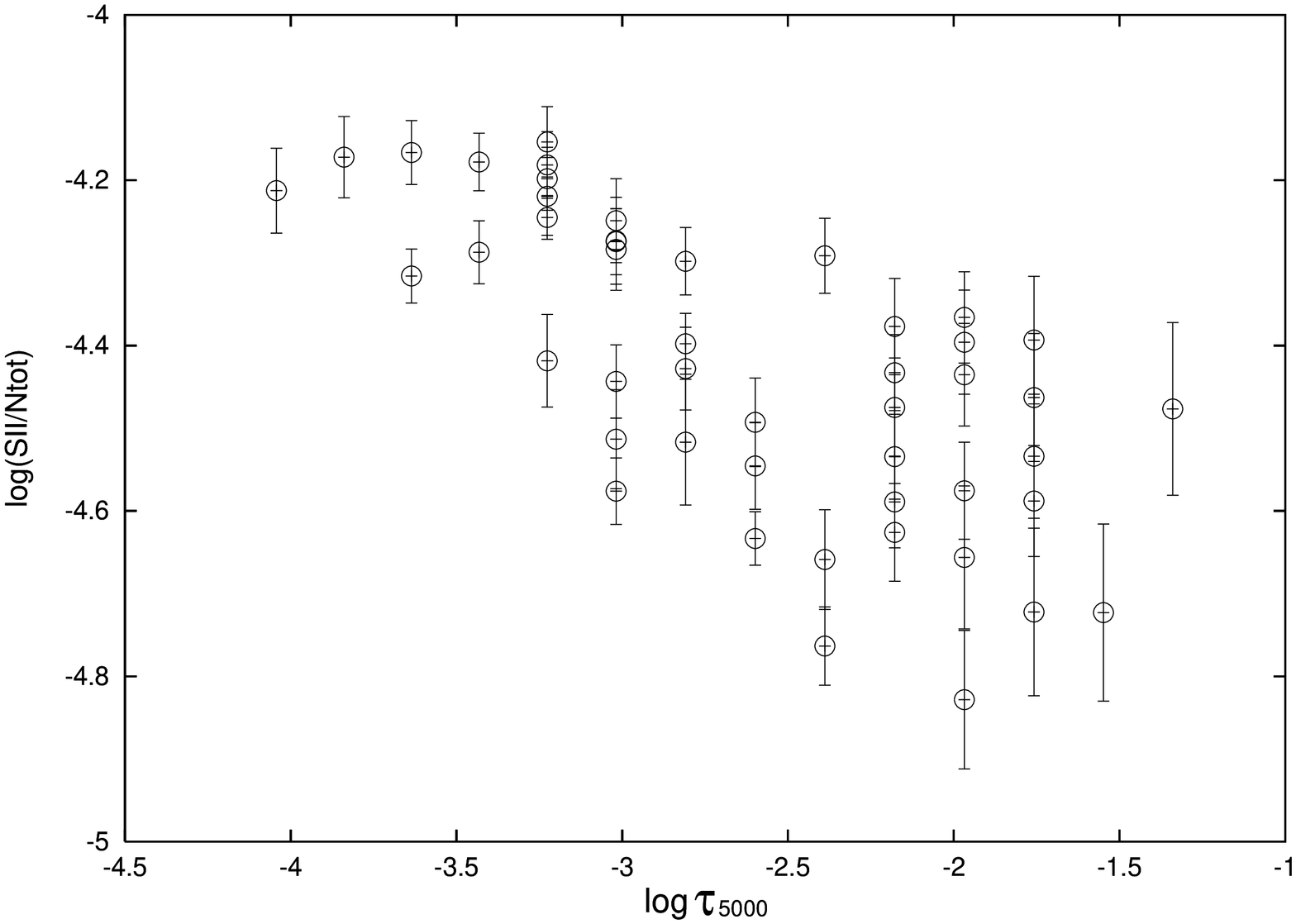}
\end{tabular}
}
\caption{ Same as in Fig.~\ref{CFe} but for a) nitrogen and b) singly ionized sulfur.
N\,{\sc i} (filled circles) and N\,{\sc ii} (open circles) lines which are formed in layers with
log $\tau_{\rm 5000}<$-1.0 require an enhanced abundance to be fit in comparison with lines
of the same ionization stage formed deeper in the atmosphere. A similar situation is found
for S\,{\sc ii} (open circles) lines formed in layers with log $\tau_{\rm 5000}<$-1.5.
\label{NS2}}
}
\end{figure*}

Carbon is enhanced by $\sim$0.9 dex in comparison with the solar abundance and
does not show strong evidence of vertical abundance stratification (see Fig.~\ref{CFe}a).
Nitrogen appears to be
strongly enhanced by $\sim$1.7 dex for the neutral species and by  $\sim$1.3 dex for the first ion.
Unlike C, it shows signatures of vertical abundance stratification (see  Fig.~\ref{NS2}a).
The abundances inferred from singly-ionized oxygen is almost the same as its solar value, while the neutral
species is enhanced by $\sim$0.5 dex.

Neon and argon appear to be enhanced respectively by $\sim$1.0 dex and
$\sim$0.8 dex in comparison with their solar abundances. The sodium abundance is estimated from the
analysis of two comparatively weak Na\,{\sc i} lines at 5688.205 \AA\, and 8194.824 \AA, which
results in a $\sim$1.6 dex enhancement.

Only one unblended Mg\,{\sc i} line (5172.684 \AA) was found, which results in a
magnesium enhancement of $\sim$0.9 dex. Meanwhile, singly ionized magnesium
is enhanced by $\sim$0.8 dex. Al\,{\sc ii} and Al\,{\sc iii} are slightly enhanced
by $\sim$0.3 dex and $\sim$0.5 dex respectively. Singly ionized silicon has only a
$\sim$0.2 dex enhancement in comparison with its solar abundance, while Si\,{\sc iii} is
enhanced by $\sim$0.6 dex. Phosphorus shows $\sim$0.9 dex enhancement.
Sulfur appears to be vertically stratified like nitrogen (see Fig.~\ref{NS2}b)
and its average atmospheric abundance is enhanced by $\sim$0.4 dex. Chlorine
seems to have a solar abundance.

\subsection{Iron-group elements: Ti to Fe}
\label{iron-group}

The spectrum of HD~135485 is extremely rich in Fe\,{\sc ii} lines.
Many of them originate from high-excitation levels with $E_{\rm i}\geq$ 10 eV.
For Fe\,{\sc ii} lines, we have used mainly Raassen~\&~Uylings (\cite{RU+98}) results to
specify their oscillator strengths and the VALD-2 database to specify the damping coefficients.
The Raassen~\&~Uylings (\cite{RU+98})
oscillator strength data usually provides a slightly better agreement
between observed and simulated Fe\,{\sc ii} line profiles than the Kurucz (\cite{Kurucz93})
GFIRON list (Pickering et al. \cite{Pickering+01}; Ryabchikova et al. \cite{Ryabchikova+05}).
The obtained average atmospheric Fe\,{\sc ii} abundance is enhanced by $\sim$0.5 dex and
{shows some signs of stratification }
(see. Fig.~\ref{CFe}b), but in an opposite manner to nitrogen and sulfur (i.e. with the concentration
{\em increasing} toward deeper layers).

Using the ESPaDOnS spectra we find that chromium and titanium
are represented by only a few unblended lines and
both elements appear to be enhanced by $\sim$0.4 dex. 
The oscillator strengths for Cr\,{\sc ii} lines
have been taken from Raassen~\&~Uylings (\cite{RU+98}), while for Ti\,{\sc ii} lines we have
used VALD-2 database.


We also checked the Sr\,{\sc ii} line at 4077.71 \AA, which is only marginally visible in these spectra.
Its abundance analysis reveals a $\sim$0.6 dex enhancement -  much less than
the value 1.24 dex enhancement obtained by Trundle et al. (\cite{Trundle+01}).
Nevertheless, the enhancements
of the other chemical species reported in Table~\ref{tab3} are in good agreement with the
results of Trundle et al. (\cite{Trundle+01}).

Adoption of different values of microturbulent velocity by Trundle et al. (\cite{Trundle+01}) and by us
should lead to systematic differences in our derived abundance. However, these differences may be reduced
because we have used different line lists and atomic data for the abundance analysis.
Trundle et al. (\cite{Trundle+01}) have used mostly NIST atomic data for all chemical species, while we have
tried to use the most precise atomic data currently available for each element (see Table~\ref{tab2}).

\begin{table*}[th]
\parbox[t]{\textwidth}{
\caption[]{Parameters of the vertical abundance distribution. Here the columns specify respectively
the element, its abundance in deeper layers, the abundance difference between the two boundaries,
the optical depth $\tau_{\rm 1}$ of the transition zone's lower
(deeper) boundary, where the element's abundance increases (or decreases) linearly, the optical depth $\tau_{\rm 2}$
of the transition zone's upper boundary, the radial velocity and $v\sin{i}$ with the respective error estimates
and the fit quality for the model with abundance stratification $\chi^\mathrm{2}_{\rm s}$
and without it $\chi^\mathrm{2}_{\rm n}$.}
\begin{tabular}{lcccccccc}
\hline
\hline
El.&\multicolumn{2}{c}{Abundance in $\log(N/N_{\rm tot})$}&
\multicolumn{2}{c}{Atmospheric depths in $\log \tau_{\rm 5000}$}& $V_{\rm r}$ & $V \sin i$& & \\
 &Low atmosphere& Abundance difference& $\log \tau_{\rm 1}$ & $\log \tau_{\rm 2}$&
km s$^{-1}$& km s$^{-1}$&$\chi^2_s$ & $\chi^2_n$\\
\hline
N & -3.15$\pm$0.03&0.68$\pm$0.25&-0.67$\pm$0.27&-2.75$\pm$0.45&-4.20$\pm$0.13&1.3$\pm$0.7&2.61& 3.83\\
S & -4.63$\pm$0.02&0.94$\pm$0.28&-1.31$\pm$0.16&-3.00$\pm$0.40&-4.24$\pm$0.08&3.0$\pm$1.2&6.87&10.06\\
\hline
\end{tabular}
\label{tab4}
}
\end{table*}

\section{Vertical abundance stratification}
\label{vert}

Applying the technique described in Sect.~\ref{proc} we have tried to determine if the abundances of
some chemical species are vertically stratified. To reach this aim, we
need to analyze as many line profiles with different
depths of formation (characterised by different
excitation potential and line strength) as possible. According to
Table~\ref{tab3}, a comprehensive analysis can only be performed for carbon, nitrogen, oxygen, sulfur
and iron, which are represented in the spectrum of HD 135485 by a sufficient number
of lines. In general, we have selected for our analysis lines
free of predicted or inferred blends.
However, if a blend is from a line of the same chemical element that forms the main line profile,
such a line was also included in our simulation.


The very sharp absorption lines of the HD~135485 (8 to 15 bins per line profile),
the relatively low signal-to-noise ratio and the uncertainties in the atomic data provide
comparatively high errors in abundances inferred from a single line.
Taking into account these uncertainties as well as ionization balance errors,
we can still see some tendencies of vertical
abundance stratification for some of the analyzed chemical elements.
Fig.~\ref{NS2} shows a systematic trend, wherein N and S lines formed
higher in the atmosphere provide abundances which are significantly larger than those
formed lower in the atmosphere. To check the influence of differences in $\log gf$ values
in different databases we have performed independent simulations of sulfur lines using NIST and VALD2
atomic data and have found similar systematic trends in both cases.

The trend observed in the abundances of N and S was
previously reported by Trundle et al. (\cite{Trundle+01}), but they interpreted it in terms
of microturbulence. However, such an interpretation would require a similar
behaviour {of the} lines of C, O and Fe, which we do not observe.
For example, although the inferred nitrogen and sulfur abundances appear to increase systematically
with line strength, carbon and oxygen abundances are approximately constant, whereas the iron abundance
shows a weak tendency to decrease with line strength (see Fig.~\ref{CFe}). The lack of a coherent
trend in the abundances derived from weak versus strong lines in spectra of different elements
is not consistent with a microturbulent interpretation, and is fully consistent with chemical stratification.
Our simulations of Fe\,{\sc ii} lines show that adoption of a microturbulent velocity $\xi=$ 2 km s$^{-1}$
results in well visible decreasing of iron abundance with line strength.
We point out that this conclusion does not necessarily imply that microturbulence is not present,
but rather that it cannot explain the observed systematic abundance trends, that its presence is not
implied by any of the data discussed in this paper, and that its presence would be surprising given
published observational and theoretical results (Landstreet 1998).
Our results therefore suggest that the observed systematic trends in the abundances of N and S imply
that these elements are vertically stratified
in the atmosphere of HD~135485.

\begin{figure}[th]
\includegraphics[angle=-90,width=4.7in]{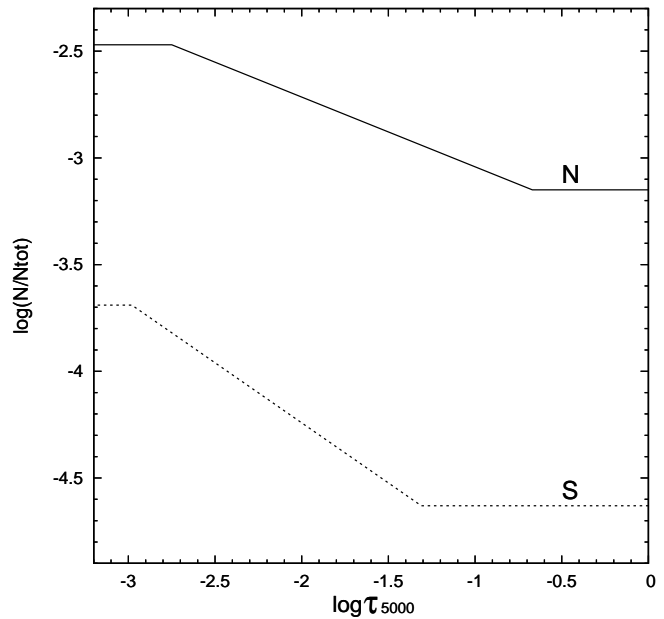}
\caption{Abundance stratification in the atmosphere of HD~135485 for nitrogen and sulfur.
\label{AbunProf}
}
\end{figure}

To obtain an estimate of the abundance distribution as a function of optical depth for N
and S, we have employed a two-zone {empirical model with a} transition zone
(Ryabchikova et al. \cite{Ryabchikova+03}; Wade et al. \cite{Wade+03}) and analyzed simultaneously
the same list of lines selected for each element. The final results for nitrogen
and sulfur stratification in the atmosphere of HD~135485 are presented in Table~\ref{tab4}
and are illustrated in Fig.~\ref{AbunProf}. They are in a good accordance with the abundance
stratification data that we have obtained from the first method for N and S (see Fig.~\ref{NS2}).
For comparison, we have also simulated the same list of lines assuming the same abundance
of the element (fitted as a free parameter) at different atmospheric depths.
From the last two columns in Table~\ref{tab4} we can
see that the model with stratified abundance distribution results in a 30\% lower $\chi^2$-function than
the model with uniform abundance. This fact shows that application of stratified model
significantly improves our fit of analysed line profiles.

The N and S abundance stratifications were also inferred assuming other stellar
atmosphere models with the same gravity $\log{g}=4.0$, but with the different effective temperatures
$T_{\rm eff}=14,500$K and $T_{\rm eff}=16,500$K, to estimate the influence of effective temperature
errors on the vertical abundance stratification profiles.
Our simulations show that an effective temperature increase does not effect the final results, while
its decrease leads to significant changes of the vertical abundance stratification profile.
The stellar atmosphere model with lower temperature results in a steeper stratification profile for
S and an increase in the average S abundance by 0.4 dex, while for nitrogen it provides
approximately a 0.8 dex increase in abundances derived from lines formed deeper in the atmosphere
where $\log\tau_{5000}>-1.0$. This leads to
a reversed stratification profile for N. Therefore the results obtained
(see Table~\ref{tab4}) for sulfur can be considered to be a lower limit for its abundance
stratification profile, whereas a uniform vertical distribution of N cannot be ruled out.



In conclusion, the analysis of our spectra shows vertical stratification of
of S and possibly N, while the elements C, O, Ne, Ar, Cr and Ti seem to have an homogeneous abundance
throughout the parts of the atmosphere that we can sample. Meanwhile,
we cannot draw firm conclusions concerning the stratification of Fe, Mg, Si,
P and Al. Some of these elements are represented by a small number (see Table~\ref{tab3})
of unblended line profiles and it is hard to obtain clear results.

\section{Discussion}
\label{discuss}

According to Trundle et al. (\cite{Trundle+01}), HD~135485 is an evolved star that probably
belongs to the BHB group. Its slow rotation, high effective temperature ($T_{\rm eff}$=15,500 K)
and observed abundance anomalies (see Table~\ref{tab3}) support the idea that microscopic
atomic diffusion is effective in its atmosphere.


Our analysis suggests that sulfur is moderately overabundant and increases in
concentration in the upper atmospheric layers. Nitrogen is strongly
overabundant and shows a similar signature of vertical stratification,
although the derived stratification of this element depends sensitively
on the adopted effective temperature (which is uncertain by about $\pm 1000$~K).
Similar results for
these two species were obtained earlier by Trundle et al. (\cite{Trundle+01}), but explained in terms
of a microturbulent velocity field. Absence of any abundance stratification for several other
elements as well as some hints that the iron abundance might be enhanced
{in} deeper atmospheric layers (i.e. opposite to the tendency shown by N and S)
shows that this interpretation is not able to
correctly explain the spectroscopic properties of this star. A more plausible
explanation is that the abundances of N and S vary as a function of
depth within the atmosphere. The detection of a similar
trend of Fe abundance versus depth in the BHB star WF4-3085 in the globular cluster M13
(Khalack et al., \cite{Khalack+06}) suggests that the trends observed in HD~135485
for N and S, and possibly Fe, are real. These facts can be considered as further arguments
in support of the efficiency of atomic diffusion in the atmosphere of
HD~135485, and possibly in the atmospheres of BHB stars generally.

The study of chemical stratification in the atmospheres of BHB stars is a recent
field of research. Previously Bonifacio et al. (\cite{Bonifacio+95}) reported
the possibility of helium stratification in the atmosphere of Feige 86.
The results obtained here for nitrogen and sulfur stratification
in HD~135485 are the first report of stratification of metal species in a
BHB star.
An extensive search for vertical stratification for a large number of chemical species in the
atmospheres of BHB stars with a range of $T_{\rm eff}$ would be
very useful to place constraints on model atmospheres in which the atmospheric
structure is calculated self-consistently with the stratification
predicted by the diffusion phenomenon, such as those of Hui-Bon-Hoa, LeBlanc
\& Hauschildt (\cite{Hui-Bon-Hoa+00}). To obtain more precise results from observations pertaining
to stratification of the elements, such models should be used since stratification
can modify the atmospheric structure and thus the inferred
stratification obtained by spectral analysis. For example, we found that the
nitrogen stratification for HD~135485 is quite sensitive to the $T_{\rm eff}$
used for the underlying model atmosphere.

\begin{acknowledgements}
This research was partially funded by the Natural Sciences and Engineering Research Council
of Canada (NSERC). We thank the R\'eseau qu\'eb\'ecois
de calcul de haute performance (RQCHP) for computing resources.
GAW acknowledges support from the
Academic Research Programme (ARP) of the Department of National Defence (Canada).
BBB thanks the staff of McDonald
Observatory for their assistance in collecting the CE data, and the
National Research Council and Naval Research Laboratory for recent
salary support.  We are grateful to
Dr. T.Ryabchikova for helpful discussions and suggestions.

\end{acknowledgements}

\end{document}